\documentclass[12pt]{article}
\usepackage{amsmath,amsthm,amssymb,euscript,epsfig,youngtab}
\usepackage{array}
\usepackage{fancybox}
\usepackage[usenames]{color}
\usepackage{amsfonts,bm}
\usepackage{graphicx}

\newcommand{\be}{\begin{equation}}
\newcommand{\ee}{\end{equation}}
\newcommand{\bea}{\begin{eqnarray}\displaystyle}
\newcommand{\eea}{\end{eqnarray}}

\makeatletter
\@addtoreset{equation}{section}
\makeatother
\renewcommand{\theequation}{\thesection.\arabic{equation}}
\def\one{{\hbox{ 1\kern-.8mm l}}}
\def\zero{{\hbox{ 0\kern-1.5mm 0}}}

\begin{document}
\makeatletter
\@addtoreset{equation}{section}
\makeatother
\renewcommand{\theequation}{\thesection.\arabic{equation}}

\rightline{WITS-CTP-132}
\vspace{1.8truecm}

\vspace{15pt}

%%%%%%%%%%%%%%%%%

{\LARGE
\centerline{\bf  Finite $N$ Quiver Gauge Theory}
}  

\vskip.5cm 

\thispagestyle{empty} \centerline{
    {\large \bf Robert de Mello Koch\footnote{{\tt robert@neo.phys.wits.ac.za}},}
   {\large \bf Rocky Kreyfelt\footnote{{\tt Rocky.Kreyfelt@students.wits.ac.za}}
    and Nkululeko Nokwara\footnote{{\tt Nkululeko.Nokwara@students.wits.ac.za}}}
                                                       }

\vspace{.4cm}
\centerline{{\it National Institute for Theoretical Physics ,}}
\centerline{{\it School of Physics and Centre for Theoretical Physics }}
\centerline{{\it University of Witwatersrand, Wits, 2050, } }
\centerline{{\it South Africa } }

\vspace{1.4truecm}

%%%%%%%%%%%%%%%%%
\thispagestyle{empty}

\centerline{\bf ABSTRACT}

\vskip.4cm 

At finite $N$ the number of restricted Schur polynomials is greater than or equal to the number of generalized restricted Schur polynomials.
In this note we study this discrepancy and explain its origin.
We conclude that, for quiver gauge theories, in general, the generalized restricted Shur polynomials correctly account for the complete set of finite $N$ 
constraints and they provide a basis, while the restricted Schur polynomials only account for a subset of the finite $N$ constraints 
and are thus overcomplete.
We identify several situations in which the restricted Schur polynomials do in fact account for the complete set of 
finite $N$ constraints.
In these situations the restricted Schur polynomials and the generalized restricted Schur polynomials both provide good bases
for the quiver gauge theory.
Finally, we demonstrate situations in which the generalized restricted Schur polynomials reduce to the restricted Schur polynomials.

\setcounter{page}{0}
\setcounter{tocdepth}{2}

\newpage

\setcounter{footnote}{0}

\linespread{1.1}
\parskip 4pt

\section{Summary and Conclusions}\label{intro}

Our focus in this article is on free gauge theories whose structure is elegantly summarized in a quiver.
By a quiver we mean a set of nodes (or vertices) connected by directed arrows, that is, a quiver is a directed graph.
The gauge group of the quiver gauge theory is a direct product of groups, one associated to each node of the quiver so that
there is a gauge field associated to each node of the quiver.
We are interested in the case that each node corresponds to a unitary group $U(N_a)$.
Although our arguments carry over to a general quiver gauge theory, we will mostly focus on quivers with two nodes,
which corresponds to studying a $U(N_1)\times U(N_2)$ gauge group.
For each directed arrow there is a bifundamental scalar.
An arrow stretching from node $a$ to node $b$ gives a field that transforms in the fundamental representation of $U(N_a)$, in the 
antifundamental of $U(N_b)$ and is a singlet of $U(N_c)$, $c\ne a,b$.

Our primary interest is in the finite $N$ physics of these theories.
A natural basis for the local gauge invariant operators of the theory is provided by taking traces of products of fields.
At finite $N$, not all trace structures are independent.
As a simple example, consider a scalar field $Z$ which is an $N\times N$ matrix transforming in the adjoint representation of $U(N)$.
A complete set of operators built using three fields is given by $\{ {\rm Tr}(Z^3),{\rm Tr}(Z^2){\rm Tr}(Z),{\rm Tr}(Z)^3\}$, when $N>2$.
For $N=2$ this set is overcomplete because we have the identity
\bea
    {\rm Tr}(Z^3)={1\over 2}\left[ 3{\rm Tr}(Z^2){\rm Tr}(Z)-{\rm Tr}(Z)^3\right]
    \label{FntN}
\eea
It is a highly non-trivial problem to write a basis of local operators that is not over complete at finite $N$.
This problem has been solved for multimatrix models with $U(N)$ gauge group in
\cite{Corley:2001zk,Kimura:2007wy,Brown:2007xh,Bhattacharyya:2008rb,bhr2,Kimura:2009wy,Kimura:2012hp,deMelloKoch:2011vn,pnm} 
and for single matrix models with $SO(N)$ or $Sp(N)$ gauge groups in \cite{Caputa:2013hr,Caputa:2013vla,Diaz:2013gja}.
The result of these studies is a basis of local operators that also diagonalizes the free field two point function.
These bases have been useful for exploring giant 
gravitons\cite{mst,myers,hash,Balasubramanian:2004nb,Berenstein:2005fa,Berenstein:2006qk,de Mello Koch:2007uu,de Mello Koch:2007uv,Bekker:2007ea,Berenstein:2013md,Berenstein:2013eya}
and new background geometries\cite{Lin:2004nb,Balasubramanian:2005mg,Berenstein:2004kk,Berenstein:2005aa,Brown:2006zk,Koch:2008ah,Koch:2008cm,deMelloKoch:2009jc,Lin:2010sba,Lin:2010nd,Kimura:2011df,Kimura:2013fqa}
in AdS/CFT\cite{Maldacena:1997re}, 
as well as for the computations of anomalous dimensions in large $N$ but non-planar 
limits\cite{Kimura:2010tx,Carlson:2011hy,Koch:2011hb,deMelloKoch:2011ci,deMelloKoch:2012ck,Koch:2013yaa,Koch:2013xaa}.
Elements in the basis are labeled by Young diagrams.
The finite $N$ relations are encoded in the statement that operators labeled by Young diagrams with more than $N$ rows vanish.
To illustrate this point note that a basis for operators built using a single field are the Schur polynomials.
For $N=2$ the constraint (\ref{FntN}) is the statement
\bea
    \chi_{\tiny \yng(1,1,1)}(Z)={1\over 6}\left( {\rm Tr}(Z)^3-3{\rm Tr}(Z^2){\rm Tr}(Z)+2{\rm Tr}(Z^3) \right) =0
\eea

For quiver gauge theories, there are two distinct approaches that have been developed to study the finite $N$ 
physics\cite{deMelloKoch:2012kv,Pasukonis:2013ts}\footnote{For earlier work, focusing on essentially single matrix 
dynamics, see \cite{Dey:2011ea,Chakrabortty:2011fd,Caputa:2012dg,Mohammed:2012xk}}. In the remainder of this introduction, we will review these 
two approaches with the goal of exhibiting a tension between them.
The primary goal of this article is to clarify the origin of this tension and to explain how it is resolved.

For concreteness, consider a quiver gauge theory with gauge group $U(N_1)\times U(N_2)$ and assume that $N_1>N_2$. 
We will use Roman indices for the $U(N_1)$ gauge group and Greek indices for the $U(N_2)$ gauge group.
Consider the problem of building gauge invariant operators using the bifundamentals $(A^I)^a_\alpha$ and $(B^I{}^\dagger)_a^\alpha$, where $I=1,2$.
It is clear that any gauge invariant operator must be a product of traces of an alternating product of $A$s and $B^\dagger$s.
This motivates the products
\bea
    \phi^{IJ}{}^a_b=(A^I)^a_\alpha (B^J{}^\dagger)_b^\alpha
\eea
which transform in the adjoint of $U(N_1)$.
Any gauge invariant single trace operator is the trace of a unique (up to cyclic permutations) product of $\phi^{IJ}$ fields.
Thus, we can use the restricted Schur polynomials\cite{Bhattacharyya:2008rb} to build a basis for the local operators of the 
quiver\cite{deMelloKoch:2012kv}.
The Young diagrams labeling these operators are cut off to have no more than $N_1$ rows.
If we had instead chosen to work with the fields
\bea
   \psi^{JI}{}^\alpha_\beta = (B^J{}^\dagger)_a^\alpha(A^I)^a_\beta
\eea
we would have constructed restricted Schur polynomials that have Young diagram labels cut off to have no more than $N_2$ rows.
These cut offs are different and they do not give the same number of gauge invariant operators, so there is a puzzle.
To see how this is resolved, restrict attention to a single field $\phi^{11}$ in which case our operators are the Schur 
polynomials $\chi_R(\phi^{11})$.
For $R\vdash d$ we obtain a Schur polynomial of degree $d$.
Recall that the degree $d$ Schur polynomials in $N$ variables are a linear basis for the space of homogeneous degree 
$d$ symmetric polynomials in $N$ variables\cite{mac}.
Thus these Schur polynomials are functions of the $N_1$ eigenvalues $\lambda_i$ of $\phi^{11}$. 
Concretely, we can write the Schur polynomial as a sum of monomials
\bea
  \chi_R(\lambda_1,\lambda_2,\ldots,\lambda_N)=\sum_T \lambda^T = \sum_T \lambda_1^{t_1}\cdots \lambda_n^{t_n}
\eea
where the summation is over all semistandard Young tableaux $T$ of shape $R$. 
The powers of the eigenvalues $t_i$ counts the number of times the number $i$ appears in $T$.
We have not yet considered the eigenvalues of
\bea
   \phi^{11}=A^1 (B^1)^\dagger
\eea
$(B^1)^\dagger$ is an $N_2\times N_1$ matrix, while $A^1$ is an $N_1\times N_2$ matrix. 
These matrices are not square, so they don't admit an eigendecomposition. 
There is however the notion of a singular value decomposition (SVD) which can be applied\cite{SVD}. 
The SVD decomposition of $(B^1)^\dagger$ is
\bea
   (B^1)^\dagger =U_B\Sigma_B V^\dagger_B
\eea
where $U_B$ is an $N_2\times N_2$ unitary matrix, $V^\dagger_B$ is an $N_1\times N_1$ unitary matrix and $\Sigma_B$ is an $N_2\times N_1$ 
rectangular matrix with non-zero singular values on its diagonal.
Since $(B^1)\dagger$ has (at most) $N_2$ non-zero singular values, the generic matrix $(B^1)^\dagger$ has a null space of dimension $N_1-N_2$. 
(Non-generic $(B^1)^\dagger$ can have an even larger null-space.) 
Of course, $\phi^{11}$ and $(B^1)^\dagger$ share the same null space, so that $\phi^{11}$ has at least $N_1-N_2$ zero eigenvalues.     

Recall that a semistandard Young tableau is column strict, that is, the entries weakly increase along each row and strictly increase down each column. 
This implies that if $R$ has more than $N_2$ rows every term in $\chi_R(\phi^{11})$ is a product of at least $N_2+1$ distinct eigenvalues. 
Since only $N_2$ of these can be non-zero, it follows that $\chi_R(\phi^{11})$ actually vanishes as soon as $R$ has more than $N_2$ rows. 
This proves that the Schur polynomials $\chi_R(\phi^{11})$ and $\chi_R(\psi^{11})$ are both cut off such that $R$ must have at most $N_2$ rows.
A very simple generalization of this reasoning allows us to conclude that we can construct restricted Schur polynomials using either $\psi^{IJ}$
or $\phi^{IJ}$.
The finite $N$ constraints are encoded in the statements that operators labeled by Young diagrams with more than\footnote{${\rm min}(N_1,N_2)$ is
equal to the smallest of $N_1$ or $N_2$.} ${\rm min}(N_1,N_2)$ rows vanish.
This implies in particular that the number of gauge invariant operators that can be constructed will depend only on the smallest of $N_1$ and $N_2$.
We will call this the restricted Schur basis.

A second approach to the finite $N$ physics entails working with the field $A^I$ and $(B^I)^\dagger$ directly\cite{Pasukonis:2013ts}.
In this case, we organize the $U(N_1)$ indices using Young diagrams that have no more than $N_1$ rows and we organize the $U(N_2)$
indices using Young diagrams that have no more than $N_2$ rows.
Thus, each operator is labeled by two types of Young diagrams that have distinct cut offs.
In this case both $N_1$ and $N_2$ enter.
This dependence is genuine and one finds, for example, that the number of operators that can be constructed depend on both $N_1$ and $N_2$.
This is the generalized restricted Schur basis\cite{Pasukonis:2013ts}.

At infinite $N$, the counting of restricted Schur polynomials and generalized restricted Schur polynomials agree. 
At finite $N$ there are more restricted Schur polynomials than there are generalized restricted Schur polynomials. 
This means that either the restricted Schur polynomials are over complete or the generalized restricted Schur polynomials are under complete. 
We will show in what follows that the restricted Schur polynomials are over complete, for a subtle reason that is peculiar to quiver gauge theories,
as we now explain. 
Given a collection of fields $\{ A^I,(B^J)^\dagger \}$we can form the fields $\phi^{IJ}$.
The number $n_{IJ}$ of each type of field is not unique and it depends on the details of how we pair the $A^I$s and the $(B^J)^\dagger$s.
To get the complete set of restricted Schur polynomials, we need to consider each possible pairing with its collections of fields described by
the numbers $\{ n_{IJ}\}$.
For a given pairing $\{ n_{IJ}\}$, the restricted Schur polynomials do give the correct finite $N$ constraints.
There are however extra genuinely new conditions that can be written which involve fields that come from different pairings,
pairing $\{ n_{IJ}\}$ and pairing $\{ n_{IJ}'\}$ say.
The restricted Schur polynomials do not respect these additional constraints and are thus over complete.
The generalized restricted Schur basis correctly accounts for the complete set of finite $N$ trace relations.
This is an important general lesson: at finite $N$ the physics of quiver gauge theories is not correctly captured by 
contracting fields to construct adjoints of specific gauge groups and then building operators from these adjoints.
The adjoints retain knowledge that they are constructed from more basic bifundamental fields in the form of extra finite $N$ relations.
To correctly account for the complete set of finite $N$ relations it seems easiest to work directly with the original bifundamental 
fields and hence the generalized restricted Schur polynomial basis.

There are exceptions to this general lesson: in certain subsectors of the theory and in specific limits, some of which we identify below, 
the restricted Schur polynomials do provide a complete basis and do account for all finite $N$ relations.
In these cases, it may be simpler to use the restricted Schur polynomials rather than the generalized restricted Schur polynomials.

In section 2 we will outline in detail, using a specific example, the origin and form of the new constraints.
There are situations in which the restricted Schur polynomials do capture the complete set of finite $N$ constraints and are 
consequently not overcomplete.
In these situations one may use either basis, as dictated by the problem being considered. 
In section 3 we will identify and describe these situations.
Section 4 considers the computation of some simple correlators which provide further useful and independent 
insight into the finite $N$ physics.
Finally in section 5 we compare the structure of the restricted Schur polynomials and the generalized restricted Schur polynomials, with
the goal of explaining why it may be simpler to use the restricted Schur polynomials rather than the generalized restricted Schur 
polynomials for certain computations.
Section 5 also demonstrates situations in which the generalized restricted Schur polynomials reduce to the restricted Schur polynomials.

In what follows we will talk of a Young diagram $r$ that has $m$ boxes or of a Young diagram $r$ that is a partition of $m$ or even more
simply, $r\vdash m$.

\section{New Finite $N$ Relations}\label{CSec}

The number of generalized restricted Schur polynomials ${\cal N}_g(n_1,n_2,m_1,m_2)$
that can be built in a theory with gauge group $U(N_1)\times U(N_2)$, using $n_1$ copies of the 
field $A^1$, $n_2$ copies of $A^2$, $m_1$ copies of $(B^1)^\dagger$ and $m_2$ copies of $(B^2)^\dagger$ is given by ($l(R)$ is the length
of the first column in $R$ and $l(S)$ is the length of the first column in $S$)\cite{Pasukonis:2013ts}
\bea
   \sum_{\small \begin{array}{c} R,S\vdash n_1+n_2\cr l(R)\le N_1\,\,\, l(S)\le N_2\end{array}}
             \sum_{\small \begin{array}{c} r_1\vdash n_1\cr r_2\vdash n_2\end{array}}
             \sum_{\small \begin{array}{c} s_1\vdash m_1\cr s_2\vdash m_2\end{array}}g(r_1,r_2,R)
             g(r_1,r_2,S)g(s_1,s_2,R)g(s_1,s_2,S)\cr
   \label{numgenrsp}
\eea
where we have $n_1+n_2=m_1+m_2$ and where $g(\cdot,\cdot,\cdot)$ is a Littlewood-Richardson coefficient.
The finite $N$ relations are accounted for by restricting the above sum so that $R$ has no more than $N_1$ rows and $S$ has no more than
$N_2$ rows.

Consider now the counting for the restricted Schur polynomial. 
The first step in the construction of the resticted Schur polynomials entails pairing the $A$s and $B^\dagger$s to produce $n_{IJ}$ copies 
of $\phi^{IJ}$. 
There is one Young diagram for each of these $\phi^{IJ}$ fields.
The number of restricted Schur polynomials is now given by ($N_-\equiv {\rm min}(N_1,N_2)$)
\bea
   {\cal N}_r(n_1,n_2,m_1,m_2)=\sum_{\{n_{IJ}\}}{\cal N}_{\{n^{IJ}\}}
   \label{numrsp}
\eea
where the above sum is a sum over all possible distinct ways of pairing, that is it is a sum over all possible distinct sets $\{ n_{IJ}\}$ 
and \cite{storm}
\bea
   {\cal N}_{\{n^{IJ}\}}=\sum_{\small \begin{array}{c} R\vdash n_1+n_2\cr l(R)\le N_-\end{array}}\sum_{r_{IJ}\vdash n_{IJ}}
                         (g(r_{11},r_{12},r_{21},r_{22};R))^2
\eea
In general, (\ref{numgenrsp}) and (\ref{numrsp}) do not agree.
The goal of this section is to explain the origin of the discrepancy.\footnote{The Littlewood-Richardson number has three indices
$g(r,s,t)$. The number $g(r,s,t)$ gives the number of times irrep $t$ of GL$_N$ appears in the tensor product of GL$_N$ representations
$r$ and $s$. By $g(r_1,r_2,...,r_n;R)$ we mean the number of times $R$ appears in the tensor product of $r_1$ with $r_2$ with $r_3$ with ... with $r_n$.
We could write this as $\sum_{s_i} g(r_1,r_2,s_1)g(s_1,r_3,s_2)\cdots g(s_{n-1},r_n,R)$.}

To make the discussion concrete, we will focus on a specific example. Consider $n_1=3$, $n_2=1$, $m_1=m_2=2$, and take $N_1,N_2 > 4$ so that
there are no finite $N$ constraints. 
In this case, a simple application of (\ref{numgenrsp}) gives ${\cal N}_g(3,1,2,2)=28$ generalized restricted Schur polynomials. 
For the number of restricted Schur polynomials, we need to consider two cases
\bea
  {\rm Case\,\, I:}\qquad  &n_{11}=2\,\,\,\, n_{12}=1\,\,\,\, n_{21}=0\,\,\,\, n_{22}=1\cr
  {\rm Case\,\, II:}\qquad &n_{11}=1\,\,\,\, n_{12}=2\,\,\,\, n_{21}=1\,\,\,\, n_{22}=0
\eea
For these cases (\ref{numrsp}) gives ${\cal N}_I$=14, ${\cal N}_{II}$=14, so that in total ${\cal N}_r(3,1,2,2)=28$.
In the next section, we prove that the number of restricted Schur polynomials and generalized restricted Schur polynomials 
always agree in the absence of finite $N$ constraints.

We will see that it is ${\cal N}_r(3,1,2,2)$ that does not correctly count the number of gauge invariant operators at finite $N$.
Since this is one of the main points of our discussion, we will give the complete details on how equation (\ref{numrsp}) is applied. 
Towards this end, we have summarized the labels for the relevant restricted Schur polynomials in Appendix \ref{LstRSP}.
Consider next the case that $N_1=N_2=2$. 
A simple application of (\ref{numgenrsp}) gives ${\cal N}_g(3,1,2,2)=13$ generalized restricted Schur polynomials.
Next, consider the complete set of possible restricted Schur polynomial labels given in Appendix \ref{LstRSP}.
For Case I, the operators given in (\ref{case11}), (\ref{case12}) and (\ref{case13}) vanish so that we have 8 operators.
For Case II, the operators given in (\ref{case21}), (\ref{case22}) and (\ref{case23}) vanish so that we have 8 operators.
This gives a total of ${\cal N}_r(3,1,2,2)=16$ restricted Schur polynomials, which shows a clear discrepancy between 
(\ref{numgenrsp}) and (\ref{numrsp}).

To explore the origin of this discrepancy, we have developed a numerical algorithm to determine the number and precise
form of the finite $N$ constraints.
Consider first the case of a single $N\times N$ matrix $Z$. 
For $N=2$ we know one of the finite $N$ constraints is given by (\ref{FntN}).
If we choose a random $2\times 2$ matrix $Z$ and form the vector
\bea
   \vec{v}=\left[\begin{array}{c} {\rm Tr}(Z^3)\cr {\rm Tr}(Z^2){\rm Tr}(Z)\cr {\rm Tr}(Z)^3\end{array}\right]
   \label{vdefn}
\eea
it will point in a random direction depending on the specific matrix $Z$.
However, we know that it must lie in a two dimensional subspace of the three dimensional space it belongs to because, thanks to
(\ref{FntN}) we know that
\bea
   \vec{v}\cdot\vec{u}=0\qquad     \vec{u}=\left[\begin{array}{c} 2\cr -3\cr 1\end{array} \right]
\eea
Now imagine preparing an ensemble of random matrices $Z^{(i)}$, $i=1,...,k$.
This ensemble of $Z^{(i)}$ can be used to construct an ensemble $\vec{v}^{(i)}$ using (\ref{vdefn}) and then we can form the matrix
\bea
   M={1\over k}\sum_{i=1}^k v^{(i)T}v^{(i)}\label{covm}
\eea
Since the $\vec{v}^{(i)}$ are all orthogonal to $\vec{u}$, but otherwise explore the orthogonal two dimensional subspace, we know that
$M$ will have a single null vector, which is $\vec{u}$ itself.

The logic clearly generalizes to multimatrix models.
We collect the complete set of multitrace structures into a vector $\vec{v}$.
By preparing an ensemble of random matrices, we can prepare an ensemble of random vectors $\vec{v}^{(i)}$ and construct the matrix $M$ as in (\ref{covm}).
Each null vector of $M$ then corresponds to a finite $N$ constraint.
In this way the finite $N$ constraints are recovered from the null vectors of $M$.
 
For Case I described above, we find a total of 14 multitrace structures are possible. 
Setting $N_1=N_2=2$ we find that $M$ has a total of 6 null vectors. 
Thus, there are 6 finite $N$ constraints leaving 8 independent multitrace operators, in perfect agreement with the number of restricted Schur polynomials.
For Case II we again find a total of 14 multitrace structures are possible and again, for $N_1=N_2=2$ we find that $M$ has 6 null vectors. 
Thus, there are 6 finite $N$ constraints leaving 8 independent multitrace operators, again in perfect agreement with the number of 
restricted Schur polynomials.
If we now form the complete set of gauge invariant operators that we can construct using $n_1=3$, $n_2=1$ and $m_1=m_2=2$, we find a total of
28 multitrace structures are possible, given by the operators of Case I and Case II above.
In this case $M$ has a total of 15 null vectors, leaving a total of 13 independent multitrace operators, in perfect agreement with the 
number of generalized restricted Schur polynomials.
At this point the origin of the discrepancy is clear.
The construction of restricted Schur polynomials starts by breaking the complete space of gauge invariant operators up into two sets,
Case I and Case II above.
By searching for the finite $N$ constraints within the operators of Case I and Case II separately, we have discovered 12 constraints.
This is 3 short of the complete set of 15 constraints discovered when searching in the complete set of gauge invariant operators.
Clearly there are some finite $N$ constraints that mix operators from Case I and operators from Case II, and these constraints are
not captured in the restricted Schur construction of \cite{deMelloKoch:2012kv}.

To summarize the conclusion of our discussion, the generalized restricted Shur polynomials correctly account for the complete set of finite $N$ 
constraints and they provide a basis, while the restricted Schur polynomials only account for a subset of the finite $N$ constraints and are 
thus overcomplete.

\section{Situations Without New Finite $N$ Relations}\label{simple}

As our discussion in the introduction suggests, in the absence of finite $N$ constraints we expect that both the generalized restricted Schur 
polynomials and the restricted Schur polynomials provide good bases.
This implies, in particular, that in the absence of finite $N$ constraints the number of restricted Schur polynomials is equal to the
number of generalized restricted Schur polynomials.
This is indeed the case as we now explain.
For concreteness we again consider a $U(N_1)\times U(N_2)$ model, building our operators from the fields $(A^I)^a_\alpha$ 
and $(B^I{}^\dagger)_a^\alpha$, where $I=1,2$.
Thus, we can form four adjoint fields $\phi^{IJ}$ and our restricted Schur polynomials are labeled by 5 Young diagrams, one 
Young diagram $r_{IJ}$ for each field $\phi^{IJ}$ and one which organizes the complete set of fields.
According to \cite{{storm},pnm} the number of restricted Schur polynomials at $N=\infty$ is given by expanding
\bea
   Z_r(t_{11},t_{12},t_{21},t_{22})&&=\sum_{n_1,n_2,m_1,m_2}\sum_{a,b,c,d}
\delta_{a+b,n_1}\delta_{c+d,n_2}\delta_{a+c,m_1}\delta_{b+d,m_2}
{\cal N}_r(n_1,n_2,m_1,m_2)t_{11}^a t_{12}^b t_{21}^c t_{22}^d\cr
   &&=\prod_{k=1}^\infty {1\over 1-t_{11}^k-t_{12}^k-t_{21}^k-t_{22}^k}
   \label{pfnumrsp}
\eea
The coefficient of $t_{11}^{n_{11}}t_{12}^{n_{12}}t_{21}^{n_{21}}t_{22}^{n_{22}}$ tells us the number of restricted Schur
polynomials that can be built using $n_{11}$ $\phi^{11}$ fields, $n_{12}$ $\phi^{12}$ fields, $n_{21}$ $\phi^{21}$ fields and $n_{22}$ 
$\phi^{22}$ fields.
The number of generalized restricted Schur polynomials at $N=\infty$ is given by expanding\cite{Pasukonis:2013ts}
\bea
   Z_g(t_{a_1},t_{a_2},t_{b_1},t_{b_2})&&=\sum_{n_1,n_2,m_1,m_2}
   {\cal N}_g(n_1,n_2,m_1,m_2)t_{a_1}^{n_1} t_{a_2}^{n_2} t_{b_1}^{m_1} t_{b_2}^{m_2}\cr
    &&=\prod_{k=1}^\infty 
    {1\over 1- (t_{a_1} t_{b_1})^k- (t_{a_1} t_{b_2})^k-(t_{a_2} t_{b_1})^k-(t_{a_2} t_{b_2})^k}
    \label{pfnumgenrsp}
\eea
The coefficient of $t_{a_1}^{n_1}t_{a_2}^{n_2}t_{b_1}^{m_1}t_{b_2}^{m_2}$ tells us how many generalized restricted Schur polynomials
can be built using $n_1$ $A_1$ fields, $n_2$ $A_2$ fields, $m_1$ $B_1^\dagger$ fields and $m_2$ $B_2^\dagger$ fields.
We can clearly transform (\ref{pfnumrsp}) into (\ref{pfnumgenrsp}) by setting $t_{ij}=t_{a_i}t_{b_j}$ which proves that
in the absence of finite $N$ constraints the number of restricted Schur polynomials is equal to the
number of generalized restricted Schur polynomials.
This change of variables provides important insight into how to relate the counting of restricted Schur polynomials and generalized 
restricted Schur polynomials, even when finite $N$ constraints play a role, as we will see.

\subsection{A single $n_{IJ}$ sector}

Consider next the case that one of $n_1,n_2,m_1,m_2$ is equal to zero.
In this case there is only one possible value for the $n_{IJ}$ so that, according to our discussion above
the restricted Schur polynomials correctly account for all finite $N$ constraints and we therefore expect the number 
of restricted Schur polynomials matches the number of generalized restricted Schur polynomials.
For concreteness, consider the case that $n_1=0$.
In this case, the Young diagram appearing in (\ref{numgenrsp}) is the Young diagram with no boxes, which we denote as $\cdot$.
Consequently,
$$
   g(r_1,r_2,R)=g(\cdot,r_2,R)=\delta_{r_2,R}\qquad g(r_1,r_2,S)=g(\cdot,r_2,S)=\delta_{r_2,S}
$$
so that the number of generalized restricted Schur polynomials (\ref{numgenrsp}) becomes
\bea
&&\sum_{\small R,S\vdash n_2\,\,\, l(R)\le N_1\,\,\, l(S)\le N_2}
             \sum_{\small  r_2\vdash n_2}
             \sum_{\small s_1\vdash m_1\,\,\, s_2\vdash m_2}\,\,\delta_{r_2,R}\,\,
             \delta_{r_2,S}g(s_1,s_2,R)g(s_1,s_2,S)\cr
  &&=\sum_{\small R \vdash n_1\,\,\, l(R)\le N_-}
             \sum_{\small s_1\vdash m_1}\sum_{\small s_2\vdash m_2}
             g(s_1,s_2,R)g(s_1,s_2,R)
\eea
To count the number of restricted Schur polynomials, note that now $r_{11}=\cdot$, $r_{12}=\cdot$, $n_{21}=m_1$ and $n_{22}=m_2$
so that (\ref{numrsp}) becomes
\bea
\sum_{\small R\vdash n_2\,\,\, l(R)\le N_-}\sum_{r_{21}\vdash m_1}\sum_{r_{22}\vdash m_{2}} (g(r_{21},r_{22};R))^2
\eea
This demonstrates an exact match between the number of restricted Schur polynomials and the number of generalized restricted Schur
polynomials as we predicted.
We will recover this result, by showing that in this case the generalized restricted Schur polynomials reduce to the restricted Schur
polynomials in section \ref{pstruct}.

\subsection{One finite rank}

Finally, consider the case that one of the ranks of the two gauge groups goes to infinity. 
For concreteness, we will take $N_2\to\infty$. 
The counting of restricted Schur polynomials is
\bea
  Z_r(t_{11},t_{12},t_{21},t_{22})=
  \sum_{\small r_{11},r_{12},r_{21},r_{22},R,l(R)\le N_1}(g(r_{11},r_{12},r_{21},r_{22};R))^2 t_{11}^{|r_{11}|} t_{12}^{|r_{12}|}
   t_{21}^{|r_{21}|} t_{22}^{|r_{22}|}
\eea
A simple change of variables gives
\bea
 Z_r =  \sum_{\small r_{11},r_{12},r_{21},r_{22},R,l(R)\le N_1}(g(r_{11},r_{12},r_{21},r_{22};R))^2
   (t_{a_1}t_{b_1})^{|r_{11}|} (t_{a_1}t_{b_2})^{|r_{12}|}
   (t_{a_2}t_{b_1})^{|r_{21}|} (t_{a_2}r_{b_2})^{|r_{22}|}\nonumber
\eea
Employing the identities
\bea
  g(r_{11},r_{12},r_{21},r_{22};R)&&= \sum_{r\vdash n_1}\sum_{s\vdash n_2}g(r_{11},r_{12},r)g(r_{21},r_{22},s)g(r,s,R)\cr
                                  &&= \sum_{t\vdash m_1}\sum_{u\vdash m_2}g(r_{11},r_{21},t)g(r_{12},r_{22},u)g(t,u,R)
\eea
we find
\bea
Z_r &&= \sum_{r,s,t,u}\sum_{R,l(R)\le N_1} g(r,s,R)g(t,u,R)
   t_{a_1}^{n_1}t_{a_2}^{n_2} t_{b_1}^{m_1}t_{b_2}^{m_2}\cr  
   &&\times \sum_{r_{11},r_{12},r_{21},r_{22}} g(r_{11},r_{12},r)g(r_{21},r_{22},s)
        g(r_{11},r_{21},t)g(r_{12},r_{22},u)
    \label{IE}
\eea
We have used $n_1=|r_{11}|+|r_{12}|$, $n_2=|r_{21}|+|r_{22}|$, $m_1=|r_{11}|+|r_{21}|$ and $m_2=|r_{12}|+|r_{22}|$ in
writing this expression.
We will now compute the sum
\bea
S=  \sum_{r_{11},r_{12},r_{21},r_{22}} g(r_{11},r_{12},r)g(r_{21},r_{22},s)
        g(r_{11},r_{21},t)g(r_{12},r_{22},u)
\eea
In the sum above, the number of rows in the $r_{IJ}$ is not restricted. 
Indeed, to capture the finite $N$ constraints, it is enough to cut the number of rows of $R$ off as we have done in (\ref{IE}).
Making use of the identity ($r\vdash n$, $s\vdash m$, $t\vdash n+m$)
\bea
  g(r,s,t)={1\over n! m!}\sum_{\sigma_1\in S_n}\sum_{\sigma_2\in S_m}\chi_r(\sigma_1)\chi_s(\sigma_2)\chi_t(\sigma_1\circ\sigma_2)
\eea
and the formula
\bea
   \sum_{R\vdash n}\chi_R(\sigma)\chi_R(\tau)=\sum_{\gamma\in S_n}\delta (\gamma\sigma\gamma^{-1}\tau^{-1})
\eea
we can write $S$ as
\bea
  S&&=\sum_{n_{i1}+n_{i2}=n_i}\sum_{n_{1i}+n_{2i}=m_i}\sum_{\psi_1\in S_{n_{11}}}\sum_{\psi_2\in S_{n_{21}}}
                            \sum_{\tau_1\in S_{n_{12}}}\sum_{\tau_2\in S_{n_{22}}}{1\over n_{11}!n_{12}!n_{21}!n_{22}!}\cr
  &&\qquad\qquad\times \chi_r(\psi_1\circ \tau_1)\chi_s(\psi_2\circ \tau_2)\chi_t(\psi_1\circ \psi_2)
                \chi_u(\tau_1\circ \tau_2)\cr
  &&=\sum_{\sigma_1\in S_{n_1}}\sum_{\sigma_2\in S_{n_2}}\sum_{\rho_1\in S_{m_1}}\sum_{\rho_2\in S_{m_2}}\sum_{\gamma\in S_{n_1+n_2}}
      {1\over n_1! n_2! m_1! m_2!}\cr
  &&\qquad\qquad\times \delta(\sigma_1\circ\sigma_2 (\rho_1\circ\rho_2)^{-1})\chi_r(\sigma_1)\chi_s(\sigma_2)\chi_t(\rho_1)\chi_u(\rho_2)\cr
  &&=\sum_{\sigma_1\in S_{n_1}}\sum_{\sigma_2\in S_{n_2}}\sum_{\rho_1\in S_{m_1}}\sum_{\rho_2\in S_{m_2}}\sum_{S\vdash n_1+n_2}
      {1\over n_1! n_2! m_1! m_2!}\cr
  &&\qquad\qquad\times \chi_S(\sigma_1\circ\sigma_2) \chi_S(\rho_1\circ\rho_2)\chi_r(\sigma_1)\chi_s(\sigma_2)\chi_t(\rho_1)\chi_u(\rho_2)\cr
  &&=\sum_{S\vdash n_1+n_2} g(r,s,S)g(t,u,S)
\eea
Plugging this back into (\ref{IE}) we find
\bea
Z_r &&= \sum_{r,s,t,u}\sum_{R,l(R)\le N_1\,\, S} g(r,s,S)g(t,u,S)g(r,s,R)g(t,u,R)
   t_{a_1}^{n_1}t_{a_2}^{n_2} t_{b_1}^{m_1}t_{b_2}^{m_2}\cr
    &&= Z_g  
\eea
proving the equality.
See Appendix \ref{Rocky} for a non-trivial example demonstrating this equality.

\section{Correlators}

In this section we will compute correlation functions of restricted Schur polynomials.
There are two things this will teach us.
First, we can confirm that the correct cut off on the number of rows of our Young diagram labels is the smallest of $N_1$ and $N_2$.
Second, we want to point out that operators from different $n_{IJ}$ sectors are not orthogonal, which corrects a statement 
in \cite{deMelloKoch:2012kv}.

The operators we study were given in\cite{deMelloKoch:2012kv}
\bea
   O_{R,\{ r\}\alpha\beta}={1\over \prod_{IJ} n_{IJ}!}
   \sum_{\sigma\in S_{n_1+n_2}} {\rm Tr}_{\{ r\}\alpha\beta}\left(\Gamma_R (\sigma)\right)
   {\rm Tr} (\sigma (\phi^{11})^{\otimes n_{11}}(\phi^{12})^{\otimes n_{12}}
   (\phi^{21})^{\otimes n_{21}}(\phi^{22})^{\otimes n_{22}})\cr
   \label{niceop}
\eea
The irrep $R$ will in general be a reducible representation of the $S_{n_{11}}\times S_{n_{12}}\times S_{n_{21}}\times S_{n_{22}}$
subgroup of $S_{n_1+n_2}$. One of the $S_{n_{11}}\times S_{n_{12}}\times S_{n_{21}}\times S_{n_{22}}$ irreps that $R$ subduces is $\{ r\}$. 
$\{r\}$ may be subduced more than once from $R$.
$\alpha$ and $\beta$ label these copies.
In the above formula, ${\rm Tr}_{\{r\}}$ is an instruction to trace only over the $\{ r\}$ subspace of the carrier space of $R$.
More precisely, we trace the row label ober the $\alpha$ copy of $\{r\}$ and the column label over the $\beta$ copy of $\{r\}$.
For simplicity we will set $n_2=0$.
The two point function
\bea
  \langle O_{R,\{ r\}\alpha\beta}O_{S,\{ s\}\gamma\delta}^\dagger \rangle = \delta_{RS}\delta_{\{r\},\{s\}}\delta_{\alpha\gamma}\delta_{\beta\delta}
     {{\rm hooks}_R f_R(N_1)f_R(N_2)\over {\rm hooks}_{r_{11}}{\rm hooks}_{r_{12}}}
   \label{corrltr}
\eea
follows immediately after using the results of \cite{deMelloKoch:2012kv}.
When the right hand side of this correlator vanishes, the operator itself vanishes.
Thus, by determining where the right hand side of this correlation function vanishes, we learn how the rows of the Young diagram labels should 
be restricted to obtain non-zero operators.
Towards this end, recall that $f_R(N)$ is a product of the factors of the Young diagram, one for each box,
where the box in row $i$ and column $j$ has factor $N-i+j$.
Consequently $f_R(N)$ vanishes whenever $R$ has more than $N$ rows.
Studying (\ref{corrltr}) we see that $R$ can have no more than $N_-$ rows where $N_-$ is the smallest of $N_1$ and $N_2$.
This is precisely the conclusion we reached in section \ref{intro}.
By studying two point functions, one can in general conclude that for gauge group
$U(N_1)\times U(N_2)\times \cdots \times U(N_p)$, all Young diagram labels must have no more than $N_-$ rows, where
$N_-$ is the smallest of $N_1,N_2,...,N_p$\cite{meup}.

To consider the case of general $n_1$, $n_2$, $m_1$, $m_2$, it proves convenient to use the operators
\bea
O_{R,\{r\}\alpha\beta}&&={\rm Tr}(P_{R,\{r\}\alpha\beta} A^{\otimes n} \,\tau\, B^{\dagger\otimes n})\cr
&&={1\over n_{11}! n_{22}!n_{12}!n_{21}!}\sum_{\sigma\in S_{n}}
{\rm Tr}_{\{r\}}\left(\Gamma_R (\sigma)\right)\prod_{i=1}^{n_1}(A_1)^{a_i}_{\alpha_{i}}\prod_{j=1+n_1}^{n}(A_2)^{a_j}_{\alpha_{j}}
(\tau)^{\alpha_1\cdots \alpha_n}_{\beta_1\cdots\beta_n}\times\cr
&& \times 
\prod_{i=1}^{n_{11}}(B_1^\dagger)^{\beta_i}_{a_{\sigma(i)}} 
\prod_{i=1+n_{11}}^{n_{1}}(B_2^\dagger)^{\beta_i}_{a_{\sigma(i)}}
\prod_{i=1+n_1}^{n_1+n_{21}}(B_1^\dagger)^{\beta_i}_{a_{\sigma(i)}} 
\prod_{i=1+n_1+n_{21}}^{n}(B_2^\dagger)^{\beta_i}_{a_{\sigma(i)}}
\label{identifyslots}
\eea
where $\tau$ is an element of the group algebra, constructed to obey
\bea
  {\rm Tr}(\tau \rho^{-1} \tau \sigma^{-1})=\delta (\rho^{-1} \sigma^{-1})
  \label{ndelta}
\eea
The two point function is\cite{deMelloKoch:2012kv}
$$
  \langle {\cal O}_{R,\{r\}\alpha\beta}{\cal O}_{S,\{s\}\gamma\delta}^\dagger \rangle =
n_{11}!n_{12}!n_{21}!n_{22}! {\rm Tr}(P_{R,\{r\}\alpha\beta} P_{S,\{s\}\gamma\delta})\, .
$$
Thus the two point function in the subspace of operators with fixed $n_{IJ}$ is diagonal.
However, even after fixing $n_I,m_J$, we can change the $n_{IJ}$. 
Projectors corresponding to different $n_{IJ}$ will not in general be orthogonal. 
The identity (\ref{ndelta}) also does not help.
Operators from different $n_{IJ}$ sectors are not orthogonal, which is again an indication that the restricted Schur basis for quiver
gauge theories is, in general, overcomplete.
Note however that the operators constructed in \cite{Pasukonis:2013ts} are a complete basis and they do diagonalize the two point function.

\section{Polynomial Structure}\label{pstruct}

The key general lesson of this article is that at finite $N$, the physics of quiver gauge theories is not correctly captured by 
contracting fields to construct adjoints of specific gauge groups. 
The fact that the adjoints are constructed from more basic bifundamental fields is reflected in extra finite $N$ relations.
To correctly account for all finite $N$ relations it seems easiest to work directly with the original bifundamental 
fields and hence the generalized restricted Schur polynomial basis. 
In section \ref{simple} we have proved that there are exceptions to this general lesson: in certain subsectors and in specific limits, the 
restricted Schur polynomials correctly account for all finite $N$ relations and hence do provide a suitable basis.
In these cases, it may be simpler to use the restricted Schur polynomials rather than the generalized restricted Schur polynomials,
as we explain in this section.
Finally, we show that when there is a single $n_{IJ}$ sector the generalized restricted Schur polynomials reduce to the
restricted Schur polynomials constructed in \cite{deMelloKoch:2012kv}.

The restricted Schur polynomial (\ref{niceop}) can be written as
\bea
   O_{R,\{ r\}\alpha\beta}={1\over \prod_{IJ} n_{IJ}!}
   \sum_{\sigma\in S_{n_1+n_2}} \sum_{a} \langle R,\{s\},\alpha,a|\Gamma_R (\sigma)|R,\{s\},\beta,a\rangle
   {\rm Tr} (\sigma (\phi^{11})^{\otimes n_{11}}(\phi^{12})^{\otimes n_{12}}
   (\phi^{21})^{\otimes n_{21}}(\phi^{22})^{\otimes n_{22}})\nonumber
\eea
Above we have explicitly written the restricted trace using the states $|R,\{s\},\gamma,a\rangle$. 
These states span a subspace of the carrier space of representation $R$ of $S_{n_1+n_2}$.
The subspace carries a representation $\{s\}$ of the subgroup $S_{n_{11}}\times S_{n_{12}}\times S_{n_{21}} \times S_{n_{22}}$.
Since $\{s\}$ will in general be subduced more than once, we need the multiplicity label $\gamma$.
Finally, index $a$ indexes states in the basis that spans the subspace.
The key technical challenge is then to develop a good enough working knowledge of the states $|R,r,\gamma,a\rangle$, that
one can carry out computations using the restricted Schur polynomials. The group theoretic quantity
\bea 
\sum_{a} \langle R,\{r\},\alpha,a|\Gamma_R (\sigma)|R,\{r\},\beta,a\rangle
\eea
is the {\it restricted character} introduced in \cite{de Mello Koch:2007uu}.

Using the same notation, the generalized restricted Schur polynomials can be written as
\bea
   O_{R,S;\{ t\},\{r\};\alpha\beta\gamma\delta}&&={1\over \prod_{IJ} n_{IJ}!}
   \sum_{\sigma,\rho\in S_{n_1+n_2}} \sum_{a,b} \langle R,\{t\},\alpha,b|\Gamma_R (\sigma)|R,\{r\},\beta,a\rangle
                                   \cr
   &&\times\langle S,\{r\},\gamma,a|\Gamma_S (\rho)|S,\{t\},\delta,b\rangle
   {\rm Tr} \left(\sigma A_{1}^{\otimes n_1}A_{2}^{\otimes n_2}\rho (B_1^\dagger)^{\otimes m_1}(B_2^\dagger)^{\otimes m_2}\right)
   \nonumber
\eea
Notice that four collections of states have been introduced: $|R,\{t\},\alpha,b\rangle$, $|R,\{r\},\beta,a\rangle$,
$|S,\{t\},\alpha,b\rangle$ and $|S,\{r\},\beta,a\rangle$. The label $\{r\}$ specifies an irrep of $S_{n_1}\times S_{n_2}$ and
$\{t\}$ specifies an irrep of $S_{m_1}\times S_{m_2}$. The collections of states introduced provide a basis for the
advertised carrier spaces, within the carrier space of $R$ and $S$, which are both irreps of $S_{n_1+n_2}$. Greek labels
are multiplicity labels. $a$ labels states within the basis of $\{r\}$ and $b$ labels states within the basis of $\{t\}$.
The group theoretic quantity
\bea
\sum_{a,b} \langle R,\{t\},\alpha,b|\Gamma_R (\sigma)|R,\{r\},\beta,a\rangle
           \langle S,\{r\},\gamma,a|\Gamma_S (\rho)|S,\{t\},\delta,b\rangle
\eea
is the {\it quiver character} introduced in \cite{Pasukonis:2013ts}.

From a group theory point of view restricted characters seem to be simpler quantities than quiver characters.
Efficient methods have been developed in \cite{Koch:2011hb} to work with restricted characters.
It remains to be seen if these methods can be extended to quiver characters.
This investigation is underway\cite{Rocky}.

Finally, consider the situation for which (say) $m_2=0$ so that there is a single $n_{IJ}$ sector.
In this case we find the generalized restricted Schur polynomial reduces to the restricted Schur polynomial
\bea
   O_{R,S;\{t\}\{S\};\alpha\delta}&&={\delta_{RS}\over \prod_{IJ} n_{IJ}!}
   \sum_{\sigma,\rho\in S_{n_1+n_2}} \sum_{a,b} \langle S,\{t\},\alpha,b|\Gamma_S (\sigma)|S,\{S\},a\rangle\cr
   && \times \langle S,\{S\},a|\Gamma_S (\rho)|S,\{t\},\delta,b\rangle
   {\rm Tr} \left(\rho A_{1}^{\otimes n_1}A_{2}^{\otimes n_2}\sigma (B_1^\dagger)^{\otimes n_1+n_2}\right)\cr
&&={\delta_{RS}\over \prod_{IJ} n_{IJ}!}
   \sum_{\sigma,\rho\in S_{n_1+n_2}} \sum_{a,b} \langle S,\{t\},\alpha,b|\Gamma_S (\sigma\rho)|S,\{t\},\delta,b\rangle\cr
   &&\times {\rm Tr} \left(\rho A_{1}^{\otimes n_1}A_{2}^{\otimes n_2}\sigma (\sigma^{-1}(B_1^\dagger)^{\otimes n_1+n_2}\sigma)\right)\cr
&&={\delta_{RS}\over \prod_{IJ} n_{IJ}!}
   \sum_{\sigma,\rho\in S_{n_1+n_2}} \sum_{a,b} \langle S,\{t\},\alpha,b|\Gamma_S (\sigma\rho)|S,\{t\},\delta,b\rangle\cr
   &&\times {\rm Tr} \left(\sigma\rho A_{1}^{\otimes n_1}A_{2}^{\otimes n_2}(B_1^\dagger)^{\otimes n_1+n_2})\right)\cr
&&= {\delta_{RS}(n_1+n_2)!\over \prod_{IJ} n_{IJ}!}
   \sum_{\sigma\in S_{n_1+n_2}} \sum_{a,b} \langle S,\{t\},\alpha,b|\Gamma_S (\sigma)|S,\{t\},\delta,b\rangle
   {\rm Tr} \left(\sigma (\phi^{11})^{\otimes n_1}(\phi^{22})^{\otimes n_2})\right)\cr
&&= {\delta_{RS}(n_1+n_2)!\over \prod_{IJ} n_{IJ}!}O_{S,\{t\},\alpha\delta}
\eea
In the above computation $\{ t\}$ specifies an irreducible representation of $S_{n_1}\times S_{n_2}$

{\vskip 0.2cm}

\noindent
{\it Acknowledgements:}
We would like to thank Sanjaye Ramgoolam for useful discussions.
This work is based upon research supported by the South African Research Chairs
Initiative of the Department of Science and Technology and National Research Foundation.
Any opinion, findings and conclusions or recommendations expressed in this material
are those of the authors and therefore the NRF and DST do not accept any liability
with regard thereto.

\begin{appendix}

\section{Restricted Schur polynomials for $n_1=3$, $n_2=1$, $m_1=m_2=2$}\label{LstRSP}

The construction of restricted Schur polynomials has been described in full generality in \cite{Bhattacharyya:2008rb}.
In this Appendix we will simply list the possible operators that can be defined.
This is all that is needed to follow the counting arguments of section \ref{CSec}.
The notation followed is to list $\chi_{R,(r_{11},r_{12},r_{21},r_{22})\alpha\beta}$ with $\alpha$ and $\beta$ multiplicity labels.
When only a single copy of representations appear there is no need for a multiplicity index and it is simply omitted.

\subsection{Case I}

\bea
 \chi_{\tiny \yng(4),(\yng(2),\yng(1),\cdot,\yng(1))}\qquad\qquad {\rm One\,\,\,operator}\label{case11}
\eea
\bea
 \chi_{\tiny \yng(3,1),(\yng(1,1),\yng(1),\cdot,\yng(1))}\qquad\qquad {\rm One\,\,\,operator}\label{case12}
\eea
\bea
 \chi_{\tiny \yng(3,1),(\yng(2),\yng(1),\cdot,\yng(1))\alpha\beta}\quad\alpha,\beta =1,2\qquad\qquad {\rm Four\,\,\,operators}\label{case13}
\eea
\bea
 \chi_{\tiny \yng(2,2),(\yng(2),\yng(1),\cdot,\yng(1))}\qquad\qquad {\rm One\,\,\,operator}
\eea
\bea
 \chi_{\tiny \yng(2,2),(\yng(2),\yng(1,1),\cdot,\yng(1))}\qquad\qquad {\rm One\,\,\,operator}
\eea
\bea
 \chi_{\tiny \yng(2,1,1),(\yng(2),\yng(1),\cdot,\yng(1))}\qquad\qquad {\rm One\,\,\,operator}
\eea
\bea
 \chi_{\tiny \yng(2,1,1),(\yng(1,1),\yng(1),\cdot,\yng(1))\alpha\beta}\quad\alpha,\beta =1,2\qquad\qquad {\rm Four\,\,\,operators}
\eea
\bea
 \chi_{\tiny \yng(1,1,1,1),(\yng(1,1),\yng(1),\cdot,\yng(1))}\qquad\qquad {\rm One\,\,\,operator}
\eea

\subsection{Case II}

\bea
 \chi_{\tiny \yng(4),(\yng(1),\yng(2),\yng(1),\cdot)}\qquad\qquad {\rm One\,\,\,operator}\label{case21}
\eea
\bea
 \chi_{\tiny \yng(3,1),(\yng(1),\yng(1,1),\yng(1),\cdot)}\qquad\qquad {\rm One\,\,\,operator}\label{case22}
\eea
\bea
 \chi_{\tiny \yng(3,1),(\yng(1),\yng(2),\yng(1),\cdot)\alpha\beta}\quad\alpha,\beta =1,2\qquad\qquad {\rm Four\,\,\,operators}\label{case23}
\eea
\bea
 \chi_{\tiny \yng(2,2),(\yng(1),\yng(2),\yng(1),\cdot)}\qquad\qquad {\rm One\,\,\,operator}
\eea
\bea
 \chi_{\tiny \yng(2,2),(\yng(1),\yng(2),\yng(1,1),\cdot)}\qquad\qquad {\rm One\,\,\,operator}
\eea
\bea
 \chi_{\tiny \yng(2,1,1),(\yng(1),\yng(2),\yng(1),\cdot)}\qquad\qquad {\rm One\,\,\,operator}
\eea
\bea
 \chi_{\tiny \yng(2,1,1),(\yng(1),\yng(1,1),\yng(1),\cdot)\alpha\beta}\quad\alpha,\beta =1,2\qquad\qquad {\rm Four\,\,\,operators}
\eea
\bea
 \chi_{\tiny \yng(1,1,1,1),(\yng(1),\yng(1,1),\yng(1),\cdot)}\qquad\qquad {\rm One\,\,\,operator}
\eea

\section{Counting when finite $N$ constraints match}\label{Rocky}

For the counting in this Appendix, we take $n_1 = 1$, $n_2 = 4$, $m_1 = 3$, $m_2 =2$, $N_1=\infty$ and $N_2=2$.
Thus, all restricted Schur polynomials labels have at most two rows. For the generalized restricted Schur polynomials,
one of the Young diagrams is unrestricted and one has at most two rows - see equation (\ref{numgenrsp}). 
In this example there are two $\left\{ n_{IJ} \right\}$ sectors of operators:
\begin{enumerate}
\item[1.] $\text{tr}\left( \sigma \phi^{11} \otimes (\phi^{21})^{\otimes 2} \otimes (\phi^{22})^{\otimes 2} \right)$
\item[2.]$\text{tr}\left( \sigma \phi^{12} \otimes (\phi^{21})^{\otimes 3} \otimes \phi^{22} \right)$
\end{enumerate}
To count the restricted Schur polynomials in sector $1$ we will use the Littlewood-Richardson numbers appearing in the following products
{\footnotesize
\bea
	\yng(2) \times \yng(2) \times \yng(1)     &&= \yng(5) + 2\yng(4,1) + 2\yng(3,2) + \yng(3,1,1) + \yng(2,2,1)\cr
	\yng(2) \times \yng(1,1) \times \yng(1)   &&= \yng(4,1) + \yng(3,2) +2\yng(3,1,1) + \yng(2,2,1) +\yng(2,1,1,1)\cr
	\yng(1,1) \times \yng(2) \times \yng(1)   &&= \yng(4,1) + \yng(3,2) +2\yng(3,1,1) + \yng(2,2,1) +\yng(2,1,1,1)\cr
	\yng(1,1) \times \yng(1,1) \times \yng(1) &&= \yng(3,2) +2\yng(2,2,1) + \yng(3,1,1) +2\yng(2,1,1,1) + \yng(1,1,1,1,1)
\eea}
To count the restricted Schur polynomials in sector $2$ we will use the Littlewood-Richardson numbers appearing in the following products
{\footnotesize
\bea
	\yng(3) \times \yng(1) \times \yng(1)     &&= \yng(5) + 2\yng(4,1) + \yng(3,2) + \yng(3,1,1)\cr
	\yng(2,1) \times \yng(1) \times \yng(1)   &&= \yng(4,1) + 2\yng(3,2) + 2\yng(2,2,1) + 2\yng(3,1,1) + \yng(2,1,1,1)\cr
	\yng(1,1,1) \times \yng(1) \times \yng(1) &&= \yng(3,1,1) + \yng(2,2,1) + 2\yng(2,1,1,1) + \yng(1,1,1,1,1)
\eea}
Restricting to Young diagrams with no more than two rows, we find
\begin{equation}
	\begin{split}
		\mathcal{N}_{l(R)\le 2} &= \mathcal{N}_1 + \mathcal{N}_2\\
								&= 14 + 11\\
								&= 25 
	\end{split}	\label{crspe}
\end{equation}

The following products appear when counting the number of generalised restricted Schur Polynomials.  
For $r_1 \vdash 1$ and $r_2 \vdash 4$
{\footnotesize
\bea
	\yng(4) \times \yng(1)      &&= \yng(5) + \yng(4,1)\cr
	\yng(3,1) \times \yng(1)    &&=  \yng(4,1) + \yng(3,2) + \yng(3,1,1)\cr
	\yng(2,2) \times \yng(1)    &&= \yng(3,2) + \yng(2,2,1)	\cr
	\yng(2,1,1)  \times \yng(1) &&= \yng(3,1,1) + \yng(2,2,1) + \yng(2,1,1,1)\cr
	\yng(1,1,1,1) \times \yng(1)&&= \yng(2,1,1,1) + \yng(1,1,1,1,1)
\eea
}
For $s_1 \vdash 3$ and $s_2 \vdash 2$
{\footnotesize
\bea
	\yng(3) \times \yng(2)     &&= \yng(5) + \yng(4,1) +\yng(3,2)\cr
	\yng(3) \times \yng(1,1)   &&= \yng(4,1) +\yng(3,1,1)\cr
	\yng(2,1) \times \yng(2)   &&= \yng(4,1) +\yng(3,2) +\yng(2,2,1) + \yng(3,1,1) \cr
	\yng(2,1) \times \yng(1,1) &&= \yng(3,2) + \yng(2,2,1) + \yng(3,1,1) +\yng(2,1,1,1)\cr
	\yng(1,1,1) \times \yng(2) &&= \yng(3,1,1) + \yng(2,1,1,1)\cr
	\yng(1,1,1) \times \yng(1,1) &&= \yng(2,2,1) +\yng(2,1,1,1) +\yng(1,1,1,1,1)
\eea
}
Using these products of Young diagrams, the number of generalised restricted Schur polynomials after restricting $l(R) \le 2 $ and 
leaving $S$ unrestricted, is $\mathcal{N} =25$ matching (\ref{crspe}).

\end{appendix}

\end{document}